\documentclass[%
 aip,
 pop,
 amsmath,amssymb,
 reprint,%
]{revtex4-1}

\usepackage{graphicx}
\usepackage{dcolumn}
\usepackage{bm}

\usepackage[utf8]{inputenc}
\usepackage[T1]{fontenc}
\usepackage{mathptmx}
\usepackage{etoolbox}

\usepackage[cal=pxtx]{mathalpha}

\newcommand{\vect}[1]{\mathbf{\boldsymbol{#1}}} 
\newcommand{\matr}[1]{\mathbf{#1}} 

\newcommand{\calD}{\mathcal{D}}
\newcommand{\calB}{\mathcal{B}}

\newcommand{\calP}{\mathcal{P}}
\newcommand{\calPm}{\mathcal{P}^{m} }

\newcommand{\Xpol}{ \vect{X}_{\text{pol}} }
\newcommand{\Bpol}{ \vect{B}_{\text{pol}} }

\newcommand{\rmd}{\mathrm{d}}

\newcommand{\hatn}{ \hat{ \vect{n} } }
\newcommand{\hats}{ \hat{ \vect{s} } }

\newcommand{\xcyc}{\vect{x}_{\text{cyc}} }

\newcommand{\calTd}{ \mathcal{ T }^{d} }
\newcommand{\calT}{ \mathcal{ T } }
\newcommand{\vectx}{ \vect{x} }
\newcommand{\vectt}{ \vect{t} }
\newcommand{\vectX}{ \vect{X} }
\newcommand{\vectm}{ \vect{m} }
\newcommand{\vectk}{ \vect{k} }
\newcommand{\vectT}{ \vect{T} }
\newcommand{\Deltatheta}[1]{ \delta^{#1}\vect{\theta} + \vectk \sast\delta^{#1} \Delta\vect\theta }

\newcommand{\bN}{ \boldsymbol{N}  }
\DeclareRobustCommand{\rchi}{{\mathpalette\irchi\relax}}
\newcommand{\irchi}[2]{\raisebox{\depth}{$#1\chi$}} 

\usepackage{mathtools} 

\usepackage{scalerel}
\usepackage{stmaryrd}

\newcommand{\sast}{\!\ast\!} 

\def\onedot{$\mathsurround0pt\ldotp$}
\def\cdddot{
  \mathbin{\vcenter{\baselineskip.67ex
    \hbox{\onedot}\hbox{\onedot}\hbox{\onedot}%
  }}%
}

\usepackage{tikz}
\usetikzlibrary{tikzmark, positioning, calc}

\tikzset{down square arrow/.style={to path={-- ++(0,-.25) -| (\tikztotarget)}}}
\tikzset{up square arrow/.style={to path={-- ++(0,+.25) -| (\tikztotarget)}}}
\tikzset{
    up square arrow cdot/.style={
        to path={-- ++(0,0.25) -| (\tikztotarget) node[pos=0.25, fill=white] {$\cdot$}}
    },
    down square arrow cdot/.style={
        to path={-- ++(0,-0.45) -| (\tikztotarget) node[pos=0.25, fill=white] {$\cdot$}}
    }
}

\usepackage{cancel}

\makeatletter
\def\@email#1#2{%
 \endgroup
 \patchcmd{\titleblock@produce}
  {\frontmatter@RRAPformat}
  {\frontmatter@RRAPformat{\produce@RRAP{*#1\href{mailto:#2}{#2}}}\frontmatter@RRAPformat}
  {}{}
}%
\makeatother
\begin{document}

\preprint{AIP/123-QED}

\title[Deformation of invariant tori]{Deformation of invariant tori under perturbation}
\author{Wenyin Wei}
     \affiliation{%
    Institute of Plasma Physics, Hefei Institutes of Physical Science, Chinese Academy of Sciences, Hefei 230031, People's Republic of China
    }
    \affiliation{
    University of Science and Technology of China, Hefei 230026, People's Republic of China
    }
    \affiliation{
    Forschungszentrum J\"{u}lich GmbH, Institute of Fusion Energy and Nuclear Waste Management – Plasma Physics, 52425 J\"{u}lich, Germany
    }
\author{Jiankun Hua}%
    \affiliation{
    Forschungszentrum J\"{u}lich GmbH, Institute of Fusion Energy and Nuclear Waste Management – Plasma Physics, 52425 J\"{u}lich, Germany
    }
    \affiliation{
    International Joint Research Laboratory of Magnetic Confinement Fusion and Plasma Physics, State Key Laboratory of Advanced Electromagnetic Engineering and Technology, School of Electrical and Electronic Engineering, Huazhong University of Science and Technology, Wuhan 430074, People’s Republic of China
    }
\author{Alexander Knieps}
    \affiliation{
    Forschungszentrum J\"{u}lich GmbH, Institute of Fusion Energy and Nuclear Waste Management – Plasma Physics, 52425 J\"{u}lich, Germany
    }

\author{Yunfeng Liang*$^{,}$}
    \email{y.liang@fz-juelich.de}
    \affiliation{%
    Institute of Plasma Physics, Hefei Institutes of Physical Science, Chinese Academy of Sciences, Hefei 230031, People's Republic of China
    }
    \affiliation{
    Forschungszentrum J\"{u}lich GmbH, Institute of Fusion Energy and Nuclear Waste Management – Plasma Physics, 52425 J\"{u}lich, Germany
    }

\date{\today}

\begin{abstract}
This study extends the functional perturbation theory~(FPT) of dynamical systems, which was initially developed for investigating the shifts of magnetic field line trajectories within the chaotic edge region of plasma when subjected to global perturbations. By contrast, invariant tori reside in the ordered regions of phase space. In magnetic confinement fusion (MCF) devices, these tori manifest as closed flux surfaces, with their nested structure governing radial transport and thus playing a critical role in confinement performance. Using the method of variation as a mathematical foundation, this Letter derives formulae that characterize the deformation of invariant tori under perturbation. These results provide new tools for targeted topology control in tokamak operations and for optimizing stellarator designs by enhancing predictive capability for flux surface behaviour.
\end{abstract}
\maketitle


\small

Invariant tori represent the ordered structures within conservative dynamical systems, such as Hamiltonian systems, in contrast to chaotic regions where long-term behavior is unpredictable. A Hamiltonian system with a higher number of integral invariants occupies more space with invariant tori, indicating a greater degree of integrability.

In MCF devices, the nested closed flux surfaces, which act as invariant tori, are crucial for achieving optimal performance. These surfaces dictate the radial transport of charged particles but can be vulnerable to disturbances like plasma responses, complex current redistribution due to plasma-wall interactions, and disruptions from the collapse of these surfaces~\cite{hudson2002, loizu2017, abdullaev2015, abdullaev2016magnetic, zhou2022, xu2023, knieps2022plasma, wei2023}.


Preserving the volume enclosed by the last closed flux surface (LCFS) is important for economic reasons in fusion reactors, given the high costs associated with the vacuum vessel's volume. Additionally, the "stickiness"~\cite{mackay1984, howard1984, harsoula2018} of the LCFS may influence the flux of particles crossing it into the scrape-off layer, acting as a finite-width separatrix when chaos arises.

Understanding the deformation of invariant tori under perturbations is crucial for predicting system behaviour and optimizing design, reducing the need for expensive or impractical real-world tests. For example, this understanding can shed light on the tight connection between the changes in magnetic topology (global structure) and local changes in the magnetic field amplitudes and directions. One example outside the MCF community is the influence of long-period comets on the stability of Solar System.

The destruction of invariant tori is a key subject in the study of the Kolmogorov-Arnold-Moser (KAM) theorem, closely tied to their deformation. As invariant tori break down, complex structures like island-around-island hierarchies and cantori with infinite gaps can emerge~\cite{mackay1984, meiss1986, alus2014, meiss2015}. This Letter lays the groundwork for further exploration of these phenomena.

This Letter adopts the standard definition of an invariant torus. For a map $\calP: \mathbb{R}^N \rightarrow \mathbb{R}^N$, if there exists a diffeomorphism $\vect\varphi: \mathbb{R}^N \supset\calTd\rightarrow\mathbb{T}^d$ (where $\mathbb{T}^d$ is the standard $d$-torus) such that the motion on $\mathbb{T}^d$ is uniformly linear but non-static—i.e., $\vect\varphi(\calP(\vectx)) - \vect\varphi(\vectx) = \Delta \vect\theta \in \mathbb{R}^d \setminus \{\vect{0}\}$ remains constant—then $\calTd$ is termed a \underline{$d$-dimensional invariant torus} (or invariant $d$-torus). For a continuous-time dynamical system, \textit{i.e.} a flow, such an invariant torus can be defined similarly. It is merely the requirement on $\vect\varphi$ becomes that the conjugate motion on 
$\calTd$ has a constant non-vanishing angular velocity $\frac{\rmd }{ \rmd t} \vect\varphi( \vectX(\vectx_0, t) )= \vect\omega \in \mathbb{R}^d \setminus \{\vect{0}\}$.  Here, $\Delta \vect\theta$ is referred to as the \textit{rotation vector} of $\calTd$, while the corresponding \textit{frequency vector} is $$\vectm := [2\pi/\Delta\theta_1, \dots, 2\pi/\Delta\theta_d].$$ 
This convention aligns with the MCF community’s practice, where field lines on a flux surface with a rotation transform $\iota=n/m=1/q$ ($n$ and $m$ being coprime) complete \underline{$m$ toroidal turns} before returning. For Poincar\'e mapping defined for one toroidal turn, $\Delta\theta \equiv 2\pi n/m \mod 2\pi$. Note that $\Delta\vect\theta$ is not unique and in the following the $\Delta\theta\in [0,2\pi)$ is adopted. An invariant torus with a commensurable rotation vector is termed $\vect\omega$-commensurable, otherwise $\vect\omega$-incommensurable. 

Let $\vect\rchi(\theta_1, \dots, \theta_d)$ be a parameterization of $\calTd$, satisfying $\calP(\vect\rchi(\theta_1,\dots, \theta_d)) = \vect\rchi(\theta_1+\Delta\theta_1, \dots, \theta_d+\Delta\theta_d)$. One can define a vector $\vectk$ as the exponent of the map $\calP$, firstly continuizing $\calP^k$ from $k\in\mathbb{Z}$ to $\mathbb{R}$, and then generalizing  to $\vectk\in\mathbb{R}^d$:
\begin{align}
\calP^\vectk \left(\vect\rchi(\vect\theta)\right) := \vect\rchi(\vect\theta + \vectk \sast \Delta\vect\theta),    
\label{eq:Pk_defining}
\end{align}
where $\sast$ denotes element-wise multiplication. Thus, $\calP^\vectm$ is, by definition, a returning map on $\calTd$. When $k \in \mathbb{R}$ instead of $\mathbb{R}^d$, $\vectk \sast \Delta \vect\theta$ reduces to $k \Delta\vect\theta$.

For the case $d = N-1$, a sequence of nested invariant $(N-1)$-tori can be parameterized as $\vect\rchi(\vect\theta, r)$, where $r$ serves as the torus label, often interpreted as the radial direction. Henceforth, we assume $d = N-1$. Researchers studying Hamiltonian systems often use $n$ torus labels, such as $\{I_i\}$, when dealing with an $n$-degree-of-freedom system with a $2N$-dimensional phase space and up to $N$ integral invariants.

However, $\vect\rchi(\vect\theta, r)$ may only be defined over a fragmented set $\mathcal{F} \subset \mathbb{R}$, reflecting the deviation of the system from integrability. In such cases, the derivative in $r$ is defined in a weaker form:
\[
\frac{\partial}{\partial r} \vect\rchi(\vect\theta, r) := \lim_{ \substack{r^\prime \rightarrow r \\ r^\prime \in \mathcal{F}} } \frac{\vect\rchi(\vect\theta, r^\prime) - \vect\rchi(\vect\theta, r)}{r^\prime - r}.
\]
Similarly, the definition of $\nabla$ is relaxed when the inverse coordinate transforms $\vect\theta(\vectx)$ and $r(\vectx)$ are only defined on fragmented subsets of $\mathbb{R}^n$. Note that $\vect\theta(\vectx)$ can be considered as a non-intersecting union of all diffeomorphisms $\vect\varphi$ corresponding to all invariant tori in the phase space.

The partial derivative $\partial_k \calP^k$ can then be easily computed by the $\vect\theta$ grid: 
\begin{align}
\partial_k \calP^k 
= \left.\dfrac{ \partial \vect\rchi }{ \partial \vect\theta }  \right|_{\mathrlap{\smash{\vect\theta + k\Delta\vect\theta}}} \cdot \Delta\vect\theta 
= \sum_{i=1}^d \Delta\theta_i~ \left.\dfrac{\partial\vect\rchi}{\partial\theta_i } \right|_{\vect\theta+k\Delta\vect\theta},    
\end{align} 
and for a vectorized $\vectk$,
\begin{align}
\partial_\vectk \calP^\vectk &=
\frac{\partial \vect\rchi }{\partial \vect\theta}
\Bigg|_{ \mathrlap{ \smash{\vect\theta+\vectk\sast\Delta\vect\theta} } } 
\cdot 
\overbrace{
    \frac{ \partial(\vectk\sast\Delta\vect\theta) }{\partial \vectk}
}^{ \mathrlap{ = \operatorname{diag}(\Delta\vect\theta)_{d\times d} } }
= \left. \begin{bmatrix}
    \mid & & \mid \\
    \Delta\theta_1 \partial_{\theta_1} \vect\rchi & \cdots & \Delta\theta_d \partial_{\theta_d} \vect\rchi \\
    \mid & & \mid 
\end{bmatrix}\right|_{ \mathrlap{\smash{
\vect\theta+\vectk\sast\Delta\vect\theta} }} 
\end{align}


As powerful mathematical tools introduced from functional analysis, partial and total functional derivatives~\cite{frigyik2008} are denoted by $\delta / \delta \mathcal{B} $ and $\rmd / \rmd \mathcal{B} $, which become directional derivatives $\Delta \mathcal{B} \cdot \delta / \delta \mathcal{B} $ and $\Delta \mathcal{B} \cdot \rmd / \rmd \mathcal{B} $ when accompanied by a given perturbation. For brevity, the former one can be simply denoted $\delta$ if the system to be perturbed and the perturbation are clear, \textit{e.g.} $\delta\calP$ and $\delta\Xpol$ denote the first variation of Poincar\'e map and the poloidal shift in standard cylindrical coordinates. In cylindrical coordinates, choose the Poicar\'e section to be an iso-$\phi$ semi-infinite section, where $\phi$ is the azimuthal angle, then the Poincar\'e map corresponds to one toroidal turn and $\delta\calP$ is simply $\delta\Xpol(\vectx_{0,\text{pol} },  \phi_s, \phi_e=\phi_s+2\pi)$ by definition. In the MCF community, equal $\Delta\theta$ for each toroidal turn ($2\pi$ change in the standard azimuthal angle $\phi$) means the angle $\theta$ is that of PEST coordinates. One can also trace the trajectory in Cartesian coordinates, let $\delta\vectX(\vectx_0,t)$ progress along the trajectory, and then transform the $\delta\vectX$ to $\delta\Xpol$ to acquire $\delta\calP$, which is simply $\delta\Xpol$ at $\phi_e= \phi_s+2\pi$. The formula~\cite{wei2024orbitshifts} describing the progression of the first variation, $\delta\vectX$, along a trajectory,
\begin{subequations}
\begin{align}
\frac{ \partial }{ \partial t } 
\delta \vect{X}
[\calB;\Delta\calB]
& ( \vect{x}_{0}, t )
=
\nabla \vect{B} \cdot \delta\vect{X}
+
\delta\vect{B} ,
\label{eq:deltaX_timeforward}
\intertext{is repeated as above and applicable in \textit{arbitrary finite-dimensional} flows. The other formula describing progression of the poloidal variation, $\delta\Xpol$, in standard three-dimensional \textit{cylindrical coordinates} is shown below}
\frac{ \partial }{ \partial \phi_e } 
\delta \Xpol
[\calB;\Delta\calB]
& ( \vect{x}_{0,\text{pol}}, \phi_s, \phi_e )
= 
\nonumber \\
&
\frac{\partial (R\Bpol / B^2_\phi) }{\partial (R,Z) } \cdot \delta\Xpol 
+
\delta \frac{R\Bpol }{B_\phi},
\label{eq:deltaXpol_phiforward}
\end{align}
\end{subequations}
where $\delta (R\Bpol / B_\phi)$ is short for $(\Delta\calB\cdot\delta / \delta \calB)  (R \Bpol / B_\phi)$, equal 
$$\frac{ R\delta\Bpol }{ B_\phi }   - \frac{ R\Bpol }{ B_\phi^2 } ~ \delta B_\phi,$$ by the product rule of differentiation.

The simple geometry analysis for the shift of a hyperbolic periodic orbit under perturbation (denoted by $\delta\xcyc$, and the toroidal turn number of the cycle denoted $\vectm$) presented in~\cite{wei2024orbitshifts},
\begin{align}
\delta\vect{x}_{\text{cyc}} 
&= - \left[
\calD\calP^\vectm - \matr{I}
\right]^{-1} \cdot \delta \mathcal{P}^\vectm ,
\label{eq:DeltaXcyc}
\end{align} 
cannot be directly extended to the case of a periodic orbit on invariant $d$-tori $\calTd$, which is the focus of this Letter. The challenge lies in the fact that $\calD\calP^\vectm - \matr{I}$ possesses eigenvalue(s) equal to zero, rendering it non-invertible. $\calD$ denotes the Jacobian matrix, which collects all the partial derivative components in $\vectx_0$. The underlying reason is that the tangent component of $\delta \vect{x}_\text{cyc}$ along $\calTd$ can take arbitrary values. To resolve this indeterminacy, it is necessary to revisit the initial geometric analysis:
\begin{align}
\delta \vect{x}_\text{cyc} 
&= \delta\mathcal{P}^\vectm (\vect{x}_\text{cyc}) 
+ \calD\calP^\vectm(\vect{x}_\text{cyc})\cdot \delta \vect{x}_\text{cyc} , 
\intertext{
 and focus solely on the normal component perpendicular to $\calTd$ by replacing $\delta \vectx_\text{cyc}$ with $\delta_\perp := \sum_{i=1}^{N-d} \hatn_i \hatn_i^\text{T} \cdot \delta $,
where $\{\hatn_i\}_{i=1}^{N-d}$ is an orthonormal basis of the local normal space $\bN_p \calTd$. Thus, the equation becomes
}
\delta_\perp \vect{x}_\text{cyc} 
&= \delta \calP^\vectm (\vect{x}_\text{cyc}) 
+ \calD\calP^\vectm(\vectx_\text{cyc})\cdot \delta_\perp \vect{x}_\text{cyc} , 
\nonumber \\
\left(
\calD\calP^\vectm - \matr{I}
\right) \cdot \delta_\perp \vect{x}_{\text{cyc}} 
&= - \delta \mathcal{P}^\vectm ,
\label{eq:to_solve_delta_perp_xcyc}
\end{align} 


However, another problem arises: $(\calD\calP^\vectm - \matr{I})\cdot \delta_\perp \vectx_\text{cyc}$ can only have a tangent component to $\calTd$ if all eigenvalues of $\calD\calP^\vectm$ are equal to one, which is usually the case for flux-preserving maps. Yet, the RHS of Eq.~\eqref{eq:to_solve_delta_perp_xcyc}, $-\delta\calP^\vectm$, depends on the perturbation and may include a normal component. There can be three possibilities:
\begin{enumerate}
    \item $\delta\calP^\vectm(\vectx_\text{cyc})$ has only a tangent component to $\calTd$.
    \item The eigenvalue(s) of $\calD\calP^\vectm$ corresponding to the eigenvectors not tangent to $\calTd$ are allowed to deviate from one, as in non-conservative systems where the divergence is non-zero. 
    \item The invariant torus is destroyed, rendering $\delta_\perp \vectx_\text{cyc}$ meaningless.
\end{enumerate}
For conservative systems, it is the first case that allows the Eq.~\eqref{eq:to_solve_delta_perp_xcyc} to hold. The tangency of $\delta\calP^\vectm(\vectx_\text{cyc})$ can be immediately acquired by imposing the functional total derivative $\Delta\calP\cdot \rmd / \rmd \calP$ on both sides of the equation defining the returning map $\calP^\vectm$,
\begin{align}
\calP^{\vectm[\calP](\vectx)} [\calP](\vectx )  
:= \vectx, &&
\\
\underbrace{ \partial_\vectm \calP^\vectm \cdot \delta \vectm }_{\mathrlap{ {= 
\sum_{i=1}^d \partial_{\theta_i}\! \vect\rchi ~\Delta\theta_i  \delta m_i
}} 
} + \delta \calP^\vectm = \vect{0},&&
\label{eq:deltaPm_tangency}
\end{align}
which clearly shows that $\delta\calP^\vectm$ must be tangent to $\calTd$ if the returning map $\calP^\vectm$ remains well-defined under perturbation. For non-conservative systems, $\delta\vectm[\calP](\vectx )$ is not well-defined because each invariant torus must be separate from others for a finite distance, meaning $\vectm(\vectx)$ is defined on a discrete subset in $\mathbb{R}^N$. Because each such torus is a sink of source in the normal directions. This subset is also dependent on $\calP$. Therefore, a partial derivative of $\vectm [\calP](\vectx)$ in $\calP$ alone is not well-defined, which has to be combined with a derivative in $\vectx$. 

Although the non-invertibility of $\calD\calP^\vectm$ is an issue for acquiring the solution of $\delta_\perp \xcyc$ directly, it can still be solved for by excluding the tangent component. One simply needs to solve for all normal components $\hatn_i^\text{T}\cdot \delta_\perp \xcyc$ by 
\begin{align}
\sum_{i=1}^{N-d}
\left(
\calD\calP^\vectm - \matr{I}
\right) \cdot \hatn_i \underline{ \hatn_i^\text{T} \cdot \delta_\perp \vect{x}_{\text{cyc}} }
&= - \delta \mathcal{P}^\vectm . 
\nonumber
\end{align}

In addition to calculating $\delta\calP^\vectm$ by progression (only applies to periodic orbits, unsuitable for quasi-periodic ones) according to Eq.~\eqref{eq:deltaXpol_phiforward}  or by $\delta m_i$ according to Eq.~\eqref{eq:deltaPm_tangency}, one can also transfer the known value of $\delta\calP^\vectm$ at a point to calculate its values at other points on the same torus. By comparing the sums to compute $\delta\calPm$ for two successive points in an $m$-periodic orbit, \textit{e.g.} $\vectx_0$ and $\vectx_1$,
\begin{align}
\delta\calPm|_{\vectx_0} =& \delta \calP|_{\vectx_{m-1}} + \calD\calP|_{\vectx_{m-1}} \cdot \delta \calP|_{\vectx_{m-2} } + \cdots \nonumber \\
&\qquad + \calD\calP^{m-1}|_{\vectx_{1}} \cdot \delta \calP|_{\vectx_{0} } , \nonumber\\
\delta\calPm|_{\vectx_1} =& \delta \calP|_{\vectx_{m}} + \calD\calP|_{\vectx_{m}} \cdot \delta \calP|_{\vectx_{m-1} } + \cdots \nonumber \\
&\qquad + \calD\calP^{m-1}|_{\vectx_{2}} \cdot \delta \calP|_{\vectx_{1} }  \nonumber\\
=& \delta \calP |_{\vectx_m} + \calD\calP|_{\vectx_m} \cdot \delta\calPm|_{\vectx_0} - \calD\calPm\cdot \delta \calP|_{\vectx_0} \nonumber \\
=& \calD\calP|_{\vectx_0} \cdot \delta\calPm|_{\vectx_0} + (\matr{I}- \calD\calPm)|_{\vectx_1} \cdot \delta \calP|_{\vectx_0} ,  \nonumber \\
\intertext{one can conclude that for successive points $\vectx_i$ and $\vectx_{i+1}$ in an $m$-periodic orbit,}
\delta\calPm|_{\vectx_{i+1}}=& \calD\calP|_{\vectx_i} \cdot \delta\calPm|_{\vectx_i} + (\matr{I}- \calD\calPm)|_{\vectx_{i+1}} \cdot \delta \calP|_{\vectx_i}
, \nonumber
\intertext{which can be generalized to the case of $\vectx\in\calTd$ as below,}
\delta\calP^\vectm|_{ \calP(\vectx) }=& \calD\calP|_{\vectx} \cdot \delta\calP^\vectm|_{\vectx} + (\matr{I}- \calD\calP^\vectm)|_{ \calP(\vectx)} \cdot \delta \calP|_{\vectx}.
\end{align}

\begin{figure}[htbp]
\includegraphics[width=1.0\linewidth]{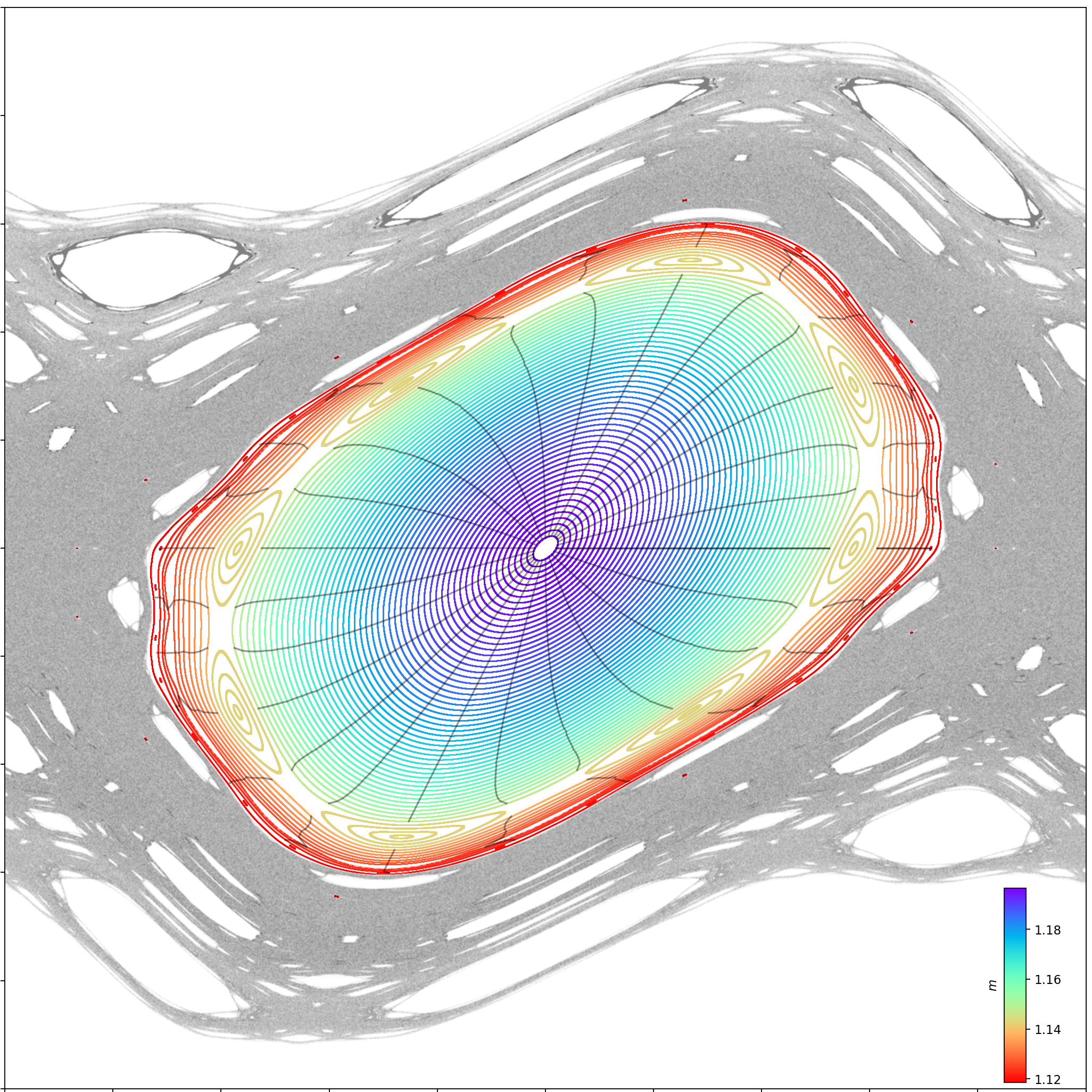}
\caption{\label{fig:Chirikov_standard_map_overview} Distribution of $m$ for the Chirikov standard map at $k=0.975$ for $\vectx\in [-0.5, 0.5]\times[-0.5, 0.5]$. Iso-$\theta$ contours are plotted with spacing $\pi/12$. }
\end{figure}
In this Letter, the standard map is taken as a demonstration of the 
present formulae with its parameter set to be $p=0.975$ (the same as Fig.~1 in \cite{meiss2015}).
To determine the exponent $m$ such that $\calPm$ is a returning map for points on an irrational invariant torus, map the initial point $\vectx_0$ for numerous times. Define an angle difference $\Delta\theta\in [0, 2\pi)$ for once mapping, then $\vectx_i$ compared to $\vectx_0$ has an angle difference $i\Delta \theta$. Denote how many times the orbit crosses $\vectx_0$ counterclockwise until the $i$-th point by $n_i$, then one knows the angle increment $i\Delta\theta$ from $\vectx_0$ to $\vectx_i$ is between $2\pi n_i \leq i\Delta\theta \leq 2\pi (n_i+1) $. Notice $m = 2\pi / \Delta \theta$, therefore
\begin{align}
\frac{i}{n_i + 1 } \leq m \leq \frac{i}{n_i}, \quad \forall i \in \mathbb{N}
\end{align}


The two variable needed to solve for $\delta_\perp \vectx_\text{cyc}$ by Eq.~\eqref{eq:to_solve_delta_perp_xcyc} are $\calD\calP^\vectm$ and $\delta\calP^\vectm$. An expression of $\calD\calP^\vectk$ in terms of $\partial_r \vect\rchi$ and $\partial_\vect\theta \vect\rchi$ is acquired by exerting total derivatives in $r$ and $\vect\theta$ \textit{resp.} on the equation~\eqref{eq:Pk_defining} defining $\calP^\vectk$,
\begin{align}
\calD\calP^\vectk \left(\vect\rchi(\vect\theta, r) \right) 
\cdot \partial_r \vect\rchi(\vect\theta, r) 
= \partial_r \vect\rchi(\vect\theta+\vectk\sast\Delta\vect\theta, r)  
  \\
+ \partial_\vect\theta \vect\rchi(\vect\theta + \vectk\sast\Delta\vect\theta, r)  \cdot
\left( \vectk \sast \frac{\rmd \Delta\vect\theta}{\rmd r}  \right)
  , \nonumber\\
\calD\calP^\vectk \left(\vect\rchi(\vect\theta, r) \right) 
\cdot \partial_\vect\theta \vect\rchi(\vect\theta, r) 
= \partial_\vect\theta \vect\rchi(\vect\theta+\vectk \sast \Delta\vect\theta, r)   
 ,
\end{align}
\begin{align}
    \calD\calP^\vectk \left(\vect\rchi(\vect\theta, r) \right) 
    = & 
    \left. \begin{bmatrix}
        \begin{array}{l}
            \partial_\vect\theta \vect\rchi \cdot (\vectk \sast\frac{\rmd \Delta\vect\theta}{\rmd r} )\\
            + \partial_r \vect\rchi 
        \end{array} &
    \partial_\vect\theta \vect\rchi
    \end{bmatrix} \right|_{ \mathrlap{\vect\theta+\vectk\sast\Delta\vect\theta} }
    \cdot
    \left.\begin{bmatrix}
        \partial_r \vect\rchi  & \partial_\vect\theta \vect\rchi  
    \end{bmatrix}\right|_{ \mathrlap{\vect\theta} }^{-1}  
\nonumber\\
    = &  \left.\begin{bmatrix}
        \mid & \mid~\mid\mid\mid~\mid \\
        \begin{array}{l} 
            \partial_\vect\theta \vect\rchi \cdot (\vectk \sast\frac{\rmd \Delta\vect\theta}{\rmd r} )  
        \end{array} & \matr{0}_{n\times (n-1)} \\
        \mid & \mid~\mid\mid\mid~\mid 
    \end{bmatrix} \right|_{ \mathrlap{\vect\theta+\vectk\sast\Delta\vect\theta} }
    \cdot
    \left.\begin{bmatrix}
        \partial_r \vect\rchi  & \partial_\vect\theta \vect\rchi  
    \end{bmatrix}\right|_{ \mathrlap{\vect\theta} }^{-1}  
\nonumber\\
    & +
    \left. \begin{bmatrix}
     | &  ||| \\
     \partial_r \vect\rchi & \partial_\vect\theta \vect\rchi \\
     | & ||| 
    \end{bmatrix} \right|_{ \mathrlap{\vect\theta+\vectk\sast\Delta\vect\theta} }
    \cdot
    \left. \begin{bmatrix}
        - ~ \nabla r ~ - \\
        \equiv \nabla \vect\theta \equiv
    \end{bmatrix} \right|_{ \mathrlap{\vect\theta} }
\end{align}
Note the inverse of $
\begin{bmatrix}
 | &  ||| \\
 \partial_r \vect\rchi & \partial_\vect\theta \vect\rchi \\
 | & ||| 
\end{bmatrix}
$ is $\begin{bmatrix}
    - ~ \nabla r ~ - \\
    \equiv \nabla \vect\theta \equiv
\end{bmatrix}$ so that 
\begin{align}
\calD\calP^\vectk ( \vect\rchi(\vect\theta, r) )
&= \left.\left(  \partial_\vect\theta \vect\rchi \cdot (\vectk\sast\frac{\rmd \Delta\vect\theta}{ \rmd r }) \right)\right|_{ \mathrlap{\smash{ \begin{subarray}{l} 
\vect\theta + \vectk \sast \Delta\vect\theta \\ n\times 1 
\end{subarray} } } }  ~~~~~ \nabla r\Big|_{ \begin{subarray}{l} \vect\theta \\ 1\times n \end{subarray} } 
\nonumber \\
& + 
\left. \begin{bmatrix}
 | &  ||| \\
 \partial_r \vect\rchi & \partial_\vect\theta \vect\rchi \\
 | & ||| 
\end{bmatrix} \right|_{ \mathrlap{\vect\theta+\vectk\sast\Delta\vect\theta} }
\cdot
\left. \begin{bmatrix}
    - ~ \nabla r ~ - \\
    \equiv \nabla \vect\theta \equiv
\end{bmatrix} \right|_{ \mathrlap{\vect\theta} },
\label{eq:DPk_by_theta_grid}
\intertext{of which a special case is when $\vectk$ takes the value of $\vectm$, }
\calD\calP^\vectm ( \vectx )
&= \left(  \partial_\vect\theta \vect\rchi \cdot (\vectm\sast\frac{\rmd \Delta\vect\theta}{ \rmd r }) \right)  \nabla r 
 +  \matr{I}_{n\times n}
\label{eq:DPm_by_theta_grid}
\end{align} where all the variables are evaluated at $\vectx$, therefore it is needless to indicate where to evaluate.

The eigenvalues of $\calD\calP^\vectm$ are identical at all points of an ($\vect\omega$-incommensurable) invariant torus. Some properties, \textit{e.g.} this one, can be transferred between an $\vect\omega$-commensurable torus and an $\vect\omega$-incommensurable torus, because in most cases the former one can be considered as a limit of a sequence of the latter ones nearby and vice versa. However, there exist extreme counterexamples in which the transfer is hindered, \textit{e.g.} an isolated invariant torus which has no invariant torus nearby. For non-conservative systems, the case of an invariant torus being isolated is not extreme but instead universal because each invariant torus is separate from others for a finite distance.

Be careful that $\calD\calP^\vectk(\vectx)$ is not $\vectm$-periodic in $\vectk$ as $\calP^\vectk$ is, which is because points on nearby invariant tori are probably to be mapped gradually away from each other as $\vectk$ increases due to the shear between tori, \textit{i.e.} neighbouring tori have a bit different rotation vectors $\Delta\vect\theta$. $\delta\calP^\vectk(\vectx)$ is also not $\vectm$-periodic but due to a reason other than shear: the impact of perturbation is accumulated all the way. 

\begin{figure}
\includegraphics[width=1.0\linewidth]{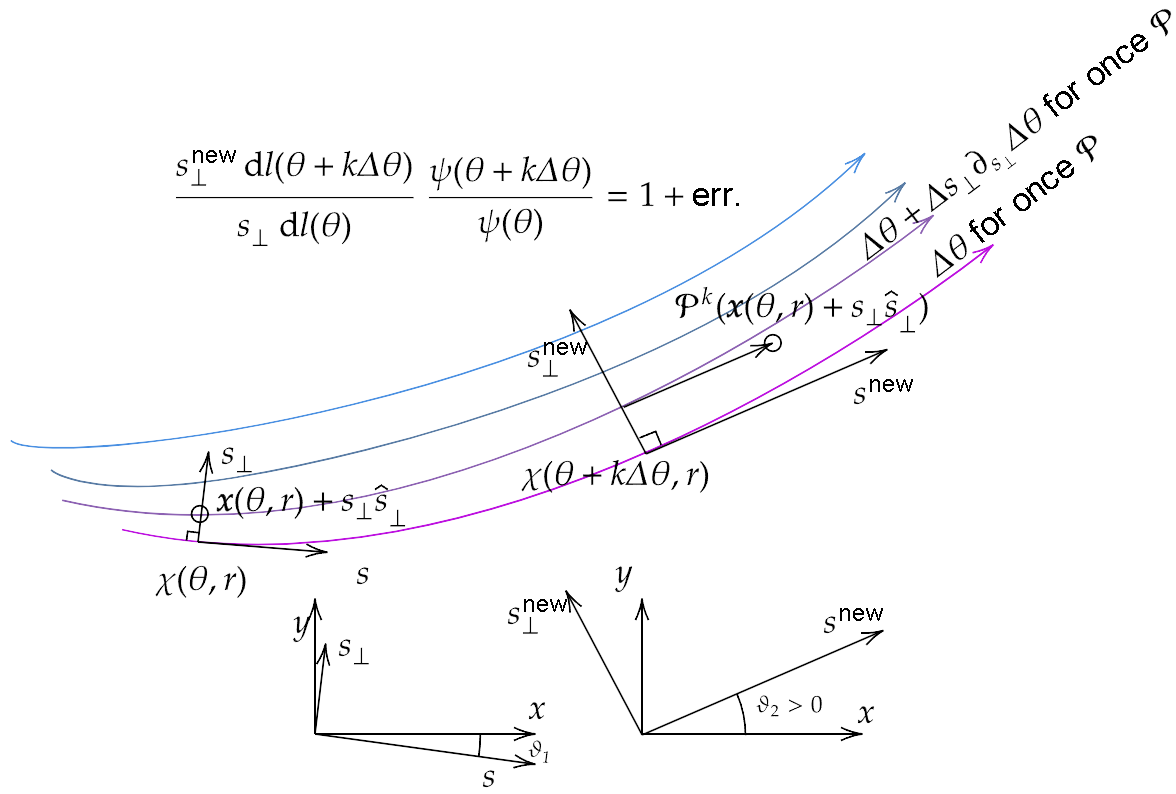}
\caption{\label{fig:DPk_flux_preserving} Cartoon to show the $\calD\calP^k$ and the local frames of coordinate systems. The point, $\vect\rchi (\theta, r) + s_\perp \hats_\perp$ , initially a bit shift form $\vect\rchi$ in the normal direction is mapped away from the $s_\perp^\text{new}$ axis due to shear.}
\end{figure}
The distance between two neighbouring invariant tori varies with which point to evaluate the distance. This is reflected by $\calD\calP^\vectm$. Let $(N,d)=(2,1)$ for illustration~(see Fig.~\ref{fig:DPk_flux_preserving}) and this case is of great importance owing to that the distance variation also reflects the local density of flux surfaces in an MCF machine. Denote a matrix representing $\varphi$ rad counterclockwise rotation in $\mathbb{R}^2$ by $\matr{R}_\varphi$. Construct local coordinate frames at $\vect{x}$ and $\calP^k(\vect{x})$ \textit{resp.} with orthonormal bases $\{\hats, \hats_\perp\}$ and $\{\hats^\text{new}, \hats_\perp^\text{new}\}$, then
\begin{align*}
\rmd \vectx_{\text{loc}}  &= \matr{R}_{-\vartheta_1} \rmd \vectx 
\nonumber\\ \\
\overbrace{
\rmd \vectx_{\text{loc}}^\text{new} 
}^{ \mathllap{ \smash{
\begin{bmatrix}
\rmd s^\text{new} \\ \rmd s_\perp^\text{new} 
\end{bmatrix} = ~~
}} }
& = \calD\calP^k_\text{loc} ~
\overbrace{ \rmd \vectx_\text{loc} }^{ \mathrlap{ \smash{
\equiv  \rmd s ~\hats + \rmd s_\perp ~\hats_\perp 
=\begin{bmatrix}
\rmd s ~ ~\\ \rmd s_\perp 
\end{bmatrix}
} } }
\nonumber\\
\rmd \vectx^\text{new}_\text{loc} 
 &= \matr{R}_{-\vartheta_2} \rmd \vectx^\text{new} 
 \nonumber\\
\rmd \vectx^\text{new} &= \underbrace{
\matr{R}_{\vartheta_2} \calD\calP^k_\text{loc} \matr{R}_{-\vartheta_1} }_{ \mathrlap{\smash{ = \calD\calP^k }} } \rmd \vectx 
\nonumber 
\end{align*}
where the subscripts $_\text{loc}$ mean the variables are viewed in the local frames. $\calD\calP^k_\text{loc}$ can be deduced from $\calD\calP^k$ or vice versa. Another relation between them is 
\begin{align}
\calD\calP^k_\text{loc} =
\begin{bmatrix}
    \hats^{\text{newT}}\cdot \calD\calP^k \cdot \hats 
    &,  & \hats^{\text{newT}}\cdot \calD\calP^k \cdot \hats_\perp \\ 
    \hats_\perp^{\text{newT}}\cdot \calD\calP^k \cdot \hats 
    &,  & \hats_\perp^{\text{newT}}\cdot \calD\calP^k \cdot \hats_\perp
\end{bmatrix}.
\end{align}

A flux-preserving property of map, \textit{e.g.} in the form of $$\psi(\vectx) \rmd S(\vectx)= \psi(\calP(\vectx)) \rmd S(\calP(\vectx)),$$(where $\psi$ is the flux density function and $\psi \rmd S$ is the flux)
gives a first-order estimation for the distance $s_\perp(\theta)$ between tori by 
\begin{align}
\psi(\theta)
s_\perp (\theta) 
\rmd l(\theta)  &= 
 \psi(\theta_0)
 s_\perp (\theta_0 ) 
    \rmd l(\theta_0)
    +..., \nonumber \\
s_\perp (\theta) &= 
0 + 
\underbrace{ s_\perp (\theta_0 ) }_{ \mathclap{ s_{\perp 0}:=\qquad } } 
\underbrace{
    \frac{\rmd l(\theta_0)}{\rmd l(\theta)}
    \frac{ \psi(\theta_0) }{ \psi(\theta) } 
    }_{ \mathclap{  
        \text{can be considered as 1st derivative }
        \frac{\rmd s_\perp  }{ \rmd s_{\perp 0} } 
    } 
    }+\mathcal{O}(|s_{\perp 0}|^2)
\nonumber \\
&= 
s_{\perp 0} 
    \frac{ |\partial_\theta \vect\rchi|_{\theta_0} }{ |\partial_\theta \vect\rchi|_\theta  }  
    \frac{\psi(\theta_0) }{\psi(\theta) }
+\mathcal{O}(|s_{\perp 0}|^2),
\end{align} where $r$ as an argument in $(\theta, r)$ is omitted for brevity since this is an expansion of $s_\perp$ near the one invariant torus of concern. For fusion devices, where the Poincaré map $\calP$ is defined for one toroidal turn, the flux density $\psi(\vectx) = B_\phi(\vectx)$.
For a general high-dimensional system, $(N,d)=(N,N-1)$, the flux-preserving property has a general form as below,
\begin{align}
\left. \det \begin{bmatrix}
    \mid & \cdots & \mid & \mid \\
    \partial_{\theta_1}\vect\rchi & \cdots & \partial_{\theta_d}\vect\rchi & s_\perp(\vect\theta) \hats_\perp \\
    \mid & \cdots & \mid & \mid 
\end{bmatrix} \right|_\vect\theta \psi(\vect\theta) =  \text{const.}, 
\end{align} (where $\hats_\perp$ is the unit vector normal to the torus)
by which one can have a similar first order estimate to $s_\perp(\vect\theta)$. 

Hereafter, our focus moves from $\calD\calP^\vectk$ to $\delta\calP^\vectk$ to provide readers with the formulae describing the deformation of invariant tori. By regarding the location $\vectx$ more fundamental than $(\vect\theta, r)$ and  considering the whole map $\calP$ also as an argument of $\vect\theta$, $\Delta\vect\theta$, $r$ and $\vect\rchi$,  the defining equation~\eqref{eq:Pk_defining} for $\calP^\vectk$ becomes
\begin{align}
\calP^\vectk [\calP] \Big( \vectx  \Big) 
:= 
\vect\rchi[\calP]\Big(\vect\theta[\calP](\vectx)+\vectk\sast\Delta
\vect\theta[\calP](\vectx),r[\calP](\vectx)\Big) ,
\label{eq:Pk_defining_fundamental_x}
\end{align} which after imposed $\Delta\calP\cdot\rmd / \rmd \calP$ converts to,
\begin{align}
& \delta\calP^\vectk (\vectx)  
=
\delta\vect\rchi(\vect\theta + \vectk\sast\Delta\vect\theta,r)  
\nonumber \\
&+ \partial_\vect\theta \vect\rchi (\vect\theta+\vectk\sast\Delta\vect\theta, r) \cdot \left(
    \delta \vect\theta (\vectx) + \vectk\sast \delta\Delta\vect\theta (\vectx) 
\right)  \nonumber\\
& + \partial_r \vect\rchi (\vect\theta + \vectk\sast\Delta\vect\theta, r) \cdot \delta r(\vectx) ,
\label{eq:Pk_on_invariantT_exerted_Ppert}
\end{align} 
that is the first form of \textit{the complete formula describing the first-order deformation of invariant tori}, allowing for any possible choice of coordinates, \textit{e.g.} where to define the curve on which $\vect\theta=\vect{0}$. For readers interested in higher-order derivations of this formula, further details are available in the Supplemental Material \cite{supp}.

When $\vectk$ takes the value of $\vectm$, one can drop $\vectk\sast\Delta\vect\theta$ because of the periodicity of $\vect\rchi$ in $\vect\theta$,
\begin{align}
\delta\calP^\vectm (\vectx) 
&=
\delta\vect\rchi(\vect\theta,r)  
+ \partial_\vect\theta \vect\rchi (\vect\theta, r) \cdot \left(
    \delta \vect\theta (\vectx) + \vectm\sast \delta\Delta\vect\theta (\vectx) 
\right)  \nonumber\\
& + \partial_r \vect\rchi (\vect\theta , r) \cdot \delta r(\vectx) ,
\label{eq:Pm_on_invariantT_exerted_Ppert}
\end{align}

\begin{figure}[htbp]
\includegraphics[width=1.0\linewidth]{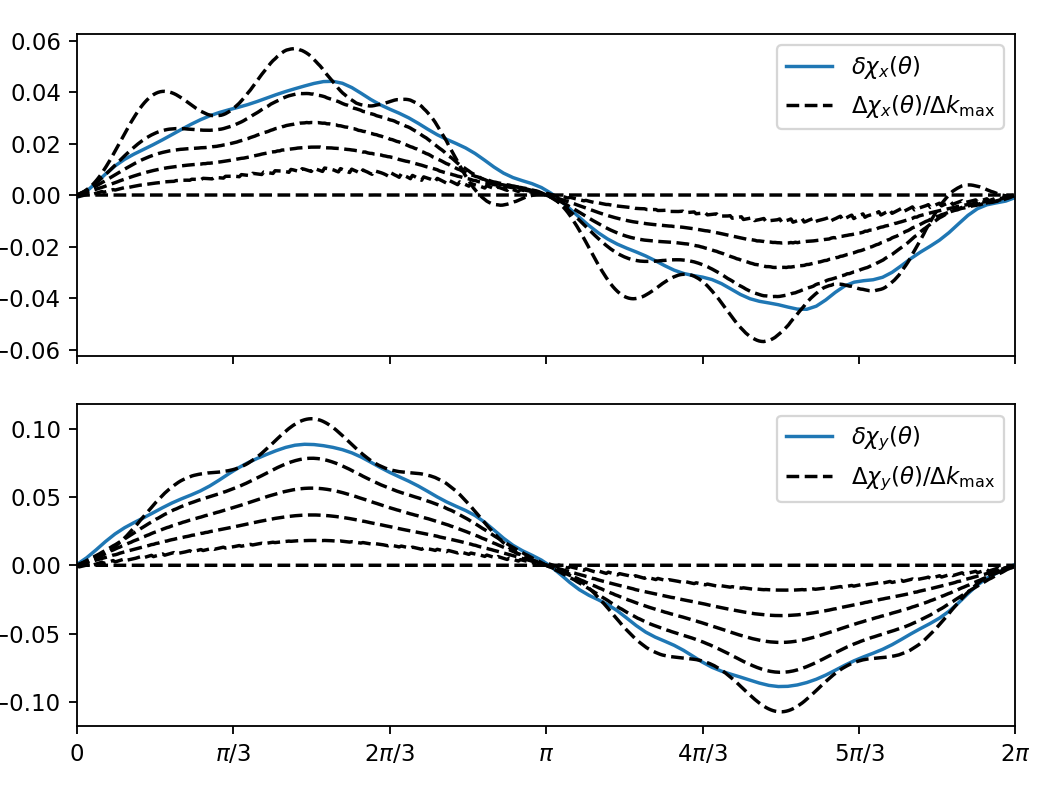}
\caption{\label{fig:Chirikov_standard_map_torus_deformation} Shifts of $\vect\rchi(\theta,r)$ for an invariant torus $\calT^1$ labelled by a starting point $\vectx_0=[.12, .0]$ at which the angle is fixed to be zero. $k_0=0.975$ while $\Delta k$ takes values of $.00, .02, .04, .06, .08, 0.10$. (a) and (b) \textit{resp.} for the $x$ and $y$ components. $\delta\vect\rchi(\theta)$ is computed here by Eq.~\eqref{eq:deformation_first_form_anchored_by_x0}. $\Delta\vect\rchi(\theta)$ is simply the difference of $\vect\rchi(\theta)$ between after and before the perturbation is imposed. }
\end{figure}
\begin{figure*}[htbp]
\includegraphics[width=1.0\linewidth]{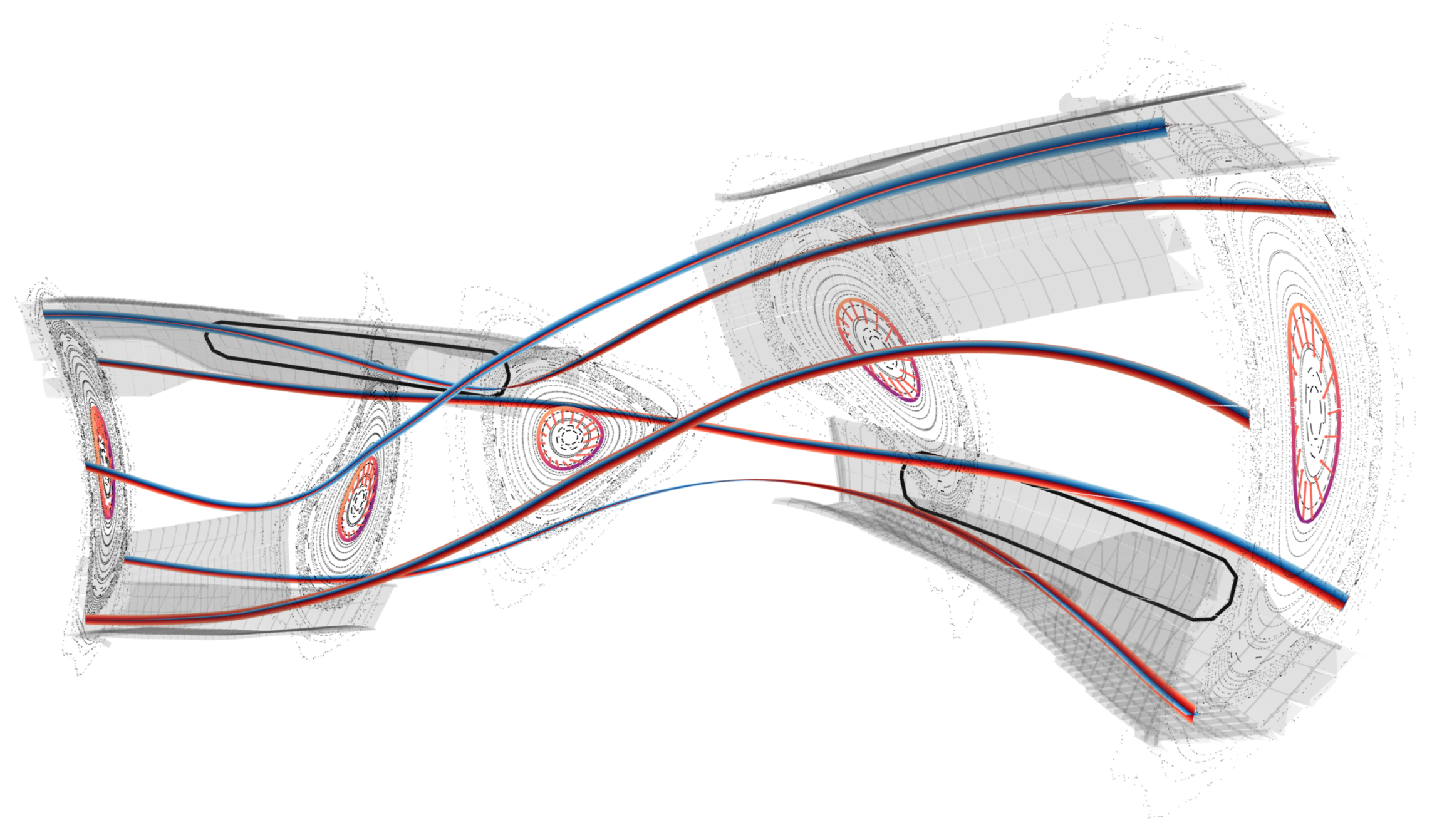}
\caption{\label{fig:w7x_flux_surface_deformation} With the perturbation field $\delta\calB$ chosen to be the vacuum field generated by the set of upper control coils (1 kAt per coil), $\delta\vect\rchi(\theta,r)$ are drawn as arrows for a flux surface ($r$ is chosen to be $\Delta\theta$, so the rotation transform is fixed during perturbation) in the standard configuration of Wendelstein 7-X, amplified by a factor of 30 for visibility. $\delta\vect\rchi(\theta, r)$ is computed here by Eq.~\eqref{eq:deformation_formula_second_form}. }
\end{figure*}
There are two common, easy-to-understand criteria to \textit{identify}, during perturbation being exerted, \textit{an invariant torus}, which can simplify the above equation by removing redundant arbitrariness in the choice of coordinates. \textcircled{1} The first one is to anchor it by a fixed point $\vectx$, which implies that $\delta r(\vectx)$ at this point always vanishes. In the meanwhile, if $\theta (\vectx)$ at this point is endowed with a constant value no matter what perturbation is imposed, $\delta\theta (\vectx)$ also vanishes. Then $\vect\rchi (\theta(\vectx), r(\vectx) )$ at this point is also fixed, \textit{i.e.} $\delta\vect\rchi(\theta,\phi)=0$. The equation~\eqref{eq:Pk_on_invariantT_exerted_Ppert} is simplified to 
\begin{align}
\delta\calP^\vectk (\vectx) 
=\delta \vect\rchi|_{\vect\theta + \vectk\sast\Delta\vect\theta} 
+ \partial_\vect\theta \vect\rchi \Big|_{
\mathrlap{\vect\theta + \vectk\sast\Delta\vect\theta
}}  \cdot 
\left( \vectk\sast \delta\Delta\vect\theta(\vectx)  \right),
\label{eq:deformation_first_form_anchored_by_x0}
\end{align}
where $\delta\calP^\vectk(\vectx) $ can be computed for $\vectk = k\sast\vect{1}$, $k\in\mathbb{Z}$, by the discrete-time version of the first variation progression Eq.~\eqref{eq:deltaXpol_phiforward}, 
\begin{align}
& \delta \calP^{k+1}(\vectx_0)
= \delta \calP(\vectx)|_{\vectx=\calP^k(\vectx_0)}
+  \delta \calP^k(\vectx_0) \cdot \calD \calP(\vectx)|_{\vectx=\calP^k(\vectx_0)}
\end{align}
while the other unknowns are $\delta\vect\rchi$~($2\pi$-periodic in every $\theta_i$) and $\delta\Delta\vect\theta(\vectx)\in \mathbb{R}^d$. One can employ least-squares methods (as used in Fig.~\ref{fig:Chirikov_standard_map_torus_deformation} is based) or other fitting techniques to estimate the Fourier series coefficients of 
$\delta\vect\rchi$ and the scalar value of 
$\delta\Delta\vect\theta(\vectx)$.

\textcircled{2} The second common choice of torus label, when $(N,d)=(2,1)$ or $(3,2)$, is \textit{resp.} $\Delta\theta$ or $\Delta\theta_1 / \Delta\theta_2$. In the former case, let $r := \Delta\theta(\vectx)$ and endow the initiating point $\vectx$ with a fixed angle $\theta$, then, Eq.~\eqref{eq:Pk_on_invariantT_exerted_Ppert} is simplified into
\begin{align}
\delta \calP^k (\vectx) = 
\delta\vect\rchi |_{\theta+k\Delta\theta} 
 + \left.\Big( 
    \partial_\theta \vect\rchi  
\Big)\right|_{\mathrlap{\smash{(\theta + k\Delta\theta,r)} } }~ \cdot \cancel{ \delta \vect\theta(\vectx) }
 +  
\left.\Big( 
    k \partial_\theta \vect\rchi  
    + \partial_r \vect\rchi 
\Big)\right|_{\mathrlap{\smash{(\theta + k\Delta\theta,r)} } }~ \cdot \delta r(\vectx)
\label{eq:deformation_for_torus_Delta_theta_is_r}
\end{align}
Owing to the fact that the perpendicular shift is an intrinsic property~(a term from differential geometry) of $\calTd$, $\delta_\perp \vectx_\text{cyc} $ computed by Eq.~\eqref{eq:to_solve_delta_perp_xcyc} shall equal the normal part of $\delta\vect\rchi(\theta, r)$ computed by Eq.~\eqref{eq:deformation_for_torus_Delta_theta_is_r}, that is $\sum_{i=1}^{N-d} \delta_\perp \vect\rchi = \hatn_i \hatn_i^\text{T} \cdot \delta \vect\rchi$, when the radial label $r$ is chosen to be $\Delta\theta$.



One may find the condition that the initiating point $\vectx$ is bound with a fixed angle not convenient, \textit{e.g.} in fusion devices, usually points at the low-field side having the identical $Z$-coordinates as that of the magnetic axis are considered to have angles $\theta=0$. To facilitate setting such a condition, consider $(\vect\theta, r)$ more fundamental than $\vectx$, \textit{i.e.} let $\vectx$ be a function of $(\vect\theta, r)$, that is $\vect\rchi(\vect\theta, r)$. Then, the defining equation~\eqref{eq:Pk_defining} for $\calP^\vectk$ becomes
\begin{align}
    \calP^\vectk [\calP]\left(
    \vect\rchi[\calP](\vect\theta,r)
    \right)
    = \vect\rchi[\calP](\vect\theta+\vectk\sast\Delta\vect\theta[\calP](r), r),
    \label{eq:Pk_defining_fundamental_theta_r}
\end{align}
which after imposed $\Delta\calP\cdot\rmd / \rmd \calP$ converts to
\begin{align}
    \delta\calP^\vectk ( \vect\rchi(\vect\theta,r))
    + \calD\calP^\vectk ( \vect\rchi(\vect\theta,r))
    \cdot  \delta\vect\rchi ( \vect\theta,r)
\nonumber \\
    =  \delta\vect\rchi (\vect\theta + \vectk\sast\Delta\vect\theta,r)
    + \left.\Big(\partial_\vect\theta \vect\rchi\right)\Big|_{\mathrlap{\smash{
        (\vect\theta + \vectk\sast\Delta\vect\theta, r) 
    } } } 
    \cdot \overbrace{
        \delta(\vectk\sast\Delta\vect\theta)
    }^{\mathclap{\text{vanishes if $\Delta\vect\theta$ is merely dependent on $r$, not on $\calP$.}}},
    \label{eq:deformation_formula_second_form}
\end{align}
that is the second complete form of \textit{the first-order deformation formula of invariant tori under perturbation}. The aforementioned condition has a natural expression 
\begin{align}
    \hat{\vect{e}}_Z \cdot \delta\vect\rchi(\theta =0, r) 
    = \text{const.} = \hat{\vect{e}}_Z \cdot \delta\vect\rchi(\theta=0, r=0).
\end{align}
For tokamaks, $\delta\vect\rchi(\theta, r)$ only needs to be solved for one iso-$\phi$ section because Poincar\'e mappings on other iso-$\phi$ sections behave the same. For stellarators, the $\delta\vect\rchi(\theta, r)$ calculated for one such section can be progressed to other sections by Eq.~\eqref{eq:deltaXpol_phiforward}, as shown in Fig.~\ref{fig:w7x_flux_surface_deformation}.

In summary, this Letter extends the functional perturbation theory of dynamical systems developed in Ref.~\onlinecite{wei2024orbitshifts, wei2024stable_manifolds_shifts} to address \textit{invariant tori and their deformation under perturbation}. This approach enables a direct connection between changes in the magnetic field and the resulting flux surface deformations, providing insight into the specific perturbations needed to achieve a desired toroidal magnetic topology. 

Notably, we do not include plasma response in this analysis but consider a general perturbation from a view of mapping to maintain the theory’s broad applicability across finite-dimensional dynamical systems emerging in various domains and to ensure clarity and accessibility for readers. Plasmas with different parameters may behave distinctly, one can use various models of plasma response magnetic field $\Delta\vect{B} = \vect{B}_\text{response} [\calB_\text{external}]+\vect{B}_\text{external}$ to calculate the resulting change of Poincar\'e map $\Delta\calP$ or merely its first variation $\delta\calP$ by Eq.~\eqref{eq:deltaXpol_phiforward}. Calligraphic font is used for the external magnetic field perturbation as an argument of $\vect{B}_\text{response} [\calB_\text{external}]$, because it is considered as a whole rather than evaluated at a specific point.

Be aware that not necessarily every point on an invariant torus has an accurate solution for $\delta_\perp \xcyc$ due to the fact that invariant tori may only be defined on a fragmented domain both in the space of $\calP$ and that of $\vectx$. An invariant torus can be destroyed into an island chain or a cantorus that has infinite gaps.


To transfer these formulae from maps to flows, one usually merely needs to replace the symbols as shown below,
\begin{align*}
\calP(\vectx_0) &\mapsto \vectX(\vectx_0, t) \text{ abbr. as }  \vectX_t (\vectx_0 ) ,
\intertext{where $t$  is fixed, so $\vectX_t$ can be considered as a map,}
\vect\varphi( \calP(\vectx) ) - \vect\varphi(\vectx) = \Delta\vect\theta  
&\mapsto \frac{\rmd \vect\varphi( \vectX(\vectx_0, t) )}{ \rmd t} = \vect\omega ,\\
\vectm := \left[
    \frac{2\pi}{\Delta\theta_1}, \dots , \frac{2\pi}{\Delta\theta_d} 
\right] &\mapsto 
\vectT := \left[
    \frac{2\pi}{\omega_1} , \dots , \frac{2\pi}{\omega_d} 
\right] ,\\
\calP^\vectk(\vect\rchi(\vect\theta)) := \vect\rchi(\vect\theta+\vectk\sast\Delta\vect\theta)
&\mapsto \vectX( \vect\rchi(\vect\theta), \vectt):=\vect\rchi(\vect\theta+\vect\omega\sast\vectt)
,\\
\calP^\vectm &\mapsto \vect{X}_{\vect{T}}.
\end{align*}


\begin{acknowledgments}
This work was supported by National Magnetic Confined Fusion Energy R\&D Program of China (No. 2022YFE03030001) and National Natural Science Foundation of China (Nos. 12275310 and 12175277). Additionally, the author Wenyin Wei would like to express his gratitude to the China Scholarship Council for providing financial support for his doctoral joint cultivation in Europe.

The authors have no conflicts to disclose. 
\end{acknowledgments}

\section*{Data Availability Statement}

The data that support the findings of this study are available from the corresponding author upon reasonable request.

\appendix*

\section{High order deformation of invariant tori}
With $\vectx$ considered more fundamental than $(\vect\theta, r)$, the defining equation~\eqref{eq:Pk_defining} for $\calP^\vectk$ is complicated into the following form by regarding the system $\calP$ as an argument of $(\calP^\vectk, \vect\rchi, \vect\theta, \Delta\vect\theta, r)$, 
\begin{align}
\calP^\vectk [\calP] \Big( \vectx  \Big) 
:= 
\vect\rchi[\calP]\Big(\vect\theta[\calP](\vectx)+\vectk\sast\Delta
\vect\theta[\calP](\vectx),r[\calP](\vectx)\Big) ,
\tag{\ref{eq:Pk_defining_fundamental_x} revisited}
\end{align} 
which after being exerted $\Delta\calP\cdot \rmd / \rmd \calP$ can yield the first form of the complete formula describing the first-order deformation of invariant tori, repeated as below, 
\begin{align}
\delta\calP^\vectk [\calP](\vectx) =
& \delta \vect\rchi + (\Deltatheta{})\cdot \partial_{\vect\theta} \vect\rchi + \delta r \cdot \partial_r \vect\rchi 
\tag{\ref{eq:Pk_on_invariantT_exerted_Ppert} revisited}
& \nonumber \\
\intertext{The more times we apply $\Delta\calP\cdot\rmd / \rmd \calP$ on both sides, the higher the order of deformation equation we obtain.}
\delta^2 \calP^\vectk [\calP](\vectx) =
& \delta^2 \vect\rchi \\
&
+ 2 (\Deltatheta{})\cdot \partial_{\vect\theta} \delta \vect\rchi 
+ 2 \delta r \cdot \partial_r \delta \vect\rchi 
\nonumber \\
& + (\Deltatheta{2})\cdot \partial_{\vect\theta} \vect\rchi 
+ (\Deltatheta{})^2 \cdot \partial_{\vect\theta}^2 \vect\rchi 
\nonumber \\
& + 2 (\Deltatheta{})\delta r \cdot \partial_r \partial_{\vect\theta} \vect\rchi  
\nonumber \\
& + \delta^2 r \cdot \partial_r \vect\rchi 
+ (\delta r)^2 \partial_r^2 \vect\rchi 
\nonumber
\end{align}

\begin{align}
    & \delta^3 \calP^\vectk [\calP](\vectx) =
    \delta^3 \vect\rchi 
\\
    & + 3 (\Deltatheta{})\cdot \partial_{\vect\theta} \delta^2 \vect\rchi 
    + 3 \delta r \cdot \partial_r \delta^2 \vect\rchi 
\nonumber\\
    & 
    + 3 (\Deltatheta{2})\cdot \partial_{\vect\theta} \delta \vect\rchi 
    + 3 (\Deltatheta{})^2 \cdot \partial_{\vect\theta}^2 \delta \vect\rchi 
\nonumber \\
    & + 6 (\Deltatheta{}) \delta r \cdot \partial_{\vect\theta} \partial_r \delta \vect\rchi 
\nonumber\\
    & + 3 (\delta r)^2  \cdot \partial_r^2 \delta \vect\rchi
    + 3 \delta^2 r \cdot \partial_r \delta\vect\rchi 
\nonumber \\
    & + (\Deltatheta{3})\cdot \partial_{\vect\theta} \vect\rchi
    + 3(\Deltatheta{})(\Deltatheta{2}) \cdot \partial_{\vect\theta}^2 \vect\rchi  
\nonumber\\
    & + (\Deltatheta{})^3 \cdot \partial_{\vect\theta}^3 \vect\rchi 
\nonumber\\
    & + 3 (\Deltatheta{2})\delta r \cdot \partial_{\vect\theta} \partial_r \vect\rchi
    + 3 (\Deltatheta{})\delta^2 r \cdot \partial_{\vect\theta} \partial_r \vect\rchi 
\nonumber \\
    & + 3 (\Deltatheta{}) (\delta r)^2  \cdot \partial_{\vect\theta} \partial_r^2 \vect\rchi
    + 3 (\Deltatheta{})^2 \delta r \cdot \partial_{\vect\theta}^2 \partial_r \vect\rchi 
\nonumber \\
    & + \delta^3 r \cdot \partial_r \vect\rchi
    + 3 \delta^2 r \delta r \cdot \partial_r^2 \vect\rchi 
\nonumber
\end{align}

One can conclude the pattern as follows (similar to that concluded for other equations in Supplemental Material of [15]), that is \textit{the first form of the high-order deformation formula of invariant torus}, (note that terms on LHS are evaluated at $\vectx$, that is $(\vect\theta, r)$, while terms on RHS are evaluated at $(\vect\theta + \vectk\sast\Delta\vect\theta, r)$)
\begin{widetext}
\begin{align}
    \frac{1}{n!} \delta^n \calP^\vectk [\calP](\vectx) 
= \sum_{\substack{ 
    \left\{ n_{\vect\theta i} \right\},
    \left\{ p_{\vect\theta i} \right\},
    \left\{ n_{r i} \right\},
    \left\{ p_{r i} \right\},
    n_{\vect\rchi} \\
    \text{such that} \\
    \sum_{i=1}^{d_{\vect\theta}}  n_{\vect\theta i} p_{\vect\theta i}
    +\sum_{i=1}^{d_{r}}  n_{r i} p_{r i}
    + n_{\vect\rchi} = n
 } } 
    \binom{ p^+_{\vect\theta} }{ p_{\vect\theta_1}, \dots, p_{\vect\theta d_\vect\theta } }
    (\frac{ \Deltatheta{n_{\vect\theta 1} } }{n_{\vect\theta 1} !} )^{p_{\vect\theta 1} }
    \cdots 
    (\frac{ \Deltatheta{n_{\vect\theta d_{\vect\theta} } } }{n_{\vect\theta d_{\vect\theta} } !} )^{p_{\vect\theta d_{\vect\theta} } } 
\nonumber \\
    \times
    \binom{ p^{+}_{r} }{ p_{r 1}, \dots, p_{r d_r } }
    ( \frac{  \delta^{n_{r 1} } r  }{ n_{r 1} ! } )^{p_{r 1}}
    ( \frac{  \delta^{n_{r d_r } } r  }{ n_{r d_r } ! } )^{p_{r d_r }}
    \cdddot_{(p^+_{\vect\theta} + p^+_r)} 
    \frac{
    \partial_{\vect\theta}^{p^+_\vect\theta} \partial_r^{p^+_r}
    \delta^{n_\vect\rchi} \vect\rchi
    }{ p^+_{\vect\theta}! p^+_{r}!  n_{\vect\rchi}! } ,
\end{align}
\end{widetext}
where 
\begin{align*} 
&  n_{\vect\rchi} \geq 0 \\
&  p^+ = p_1 + p_2 + \cdots + p_d, \quad d \text{ is the total number of powers,}\\
&  p_i \geq 1, \\ 
&  n_1 > n_2 > \cdots > n_d \geq 1,
\end{align*}
with subscripts $_\vect\theta$ and $_r$ of $(n_{\vect\theta i}, p_{\vect\theta i}, p^+_{\vect\theta}, d_{\vect\theta}, n_{r ~i}, p_{r~ i}, p^+_{r}, d_{r})$ omitted for brevity. The formula can be reduced by removing the redundant arbitrariness in choosing coordinates.

On the other hand, with $(\vect\theta, r)$ considered more fundamental than $\vectx$, the defining equation~\eqref{eq:Pk_defining} for $\calP^\vectk$ is complicated into the following form by regarding the system $\calP$ as an argument of $\calP^\vectk$, $\vect\rchi$ and $\Delta\vect\theta$, 
\begin{align}
    \calP^\vectk [\calP]\left(
    \vect\rchi[\calP](\vect\theta,r)
    \right)
    = \vect\rchi[\calP](\vect\theta+\vectk\sast\Delta\vect\theta[\calP](r), r),
    \tag{\ref{eq:Pk_defining_fundamental_theta_r} revisited}
\end{align}
which after imposed $\Delta\calP\cdot\rmd / \rmd \calP$ converts to
\begin{align}
    \delta\calP^\vectk ( \vect\rchi(\vect\theta,r))
    + \calD\calP^\vectk ( \vect\rchi(\vect\theta,r))
    \cdot  \delta\vect\rchi ( \vect\theta,r)
\nonumber \\
    =  \delta\vect\rchi (\vect\theta + \vectk\sast\Delta\vect\theta,r)
    + \left.\Big(\partial_\vect\theta \vect\rchi\right)\Big|_{\mathrlap{\smash{
        (\vect\theta + \vectk\sast\Delta\vect\theta, r) 
    } } } 
    \cdot 
        \delta(\vectk\sast\Delta\vect\theta)
    ,
    \tag{\ref{eq:deformation_formula_second_form} revisited}
\intertext{with arguments omitted for brevity,}
    \delta\calP^\vectk
    + \delta\vect\rchi  \cdot \calD\calP^\vectk
    =  \delta\vect\rchi
    + (\vectk\sast\delta\Delta\vect\theta) \cdot \partial_\vect\theta \vect\rchi .
\end{align}
The second-order formula is 
\begin{align}
    \delta^2 \calP^\vectk
    + 2 (\delta\vect\rchi \cdot \calD) \delta \calP^\vectk
    +   (\delta^2 \vect\rchi \cdot \calD)  \calP^\vectk
    +   (\delta   \vect\rchi \cdot \calD)^2\calP^\vectk
\nonumber\\
    =  \delta^2 \vect\rchi
    + 2 (\vectk\sast\delta\Delta\vect\theta) \cdot \partial_\vect\theta \delta \vect\rchi 
    +   (\vectk\sast\delta  \Delta\vect\theta)^2 \cdot \partial^2_\vect\theta \vect\rchi 
    +   (\vectk\sast\delta^2 \Delta\vect\theta) \cdot \partial_\vect\theta \delta \vect\rchi 
\end{align}
Hence, \textit{the second form of the high-order deformation formula of invariant torus} is concluded as bellow with denotations defined in a similar manner as those of the first form, (note that terms on LHS are evaluated at $(\vect\theta, r)$, while terms on RHS are evaluated at $(\vect\theta + \vectk\sast\Delta\vect\theta, r)$)
\begin{widetext}
\begin{align}
\sum_{\mathclap{\substack{ \\ \\
        \left\{ n_{\vect\rchi i} \right\},
        \left\{ p_{\vect\rchi i} \right\},
        n_{\calP} 
        \text{ such that} \\
        \sum_{i=1}^{d_{\vect\rchi}}  n_{\vect\rchi i} p_{\vect\rchi i}
        + n_{\calP} = n
     } } }
    \binom{ p^+_{\vect\rchi} }{ p_{\vect\rchi_1}, \dots, p_{\vect\rchi d_\vect\rchi } }
    (\frac{ \delta^{n_{\vect\rchi 1}} \vect\rchi  }{n_{\vect\rchi 1} !} )^{p_{\vect\rchi 1} }
    \cdots 
    (\frac{ \delta^{n_{\vect\rchi d_\vect\rchi }} \vect\rchi }{n_{\vect\rchi d_\vect\rchi} !} )^{p_{\vect\rchi d_{\vect\rchi} } } 
    \cdddot_{(p^+_{\vect\rchi})} 
    \frac{
        \calD^{p^+_\vect\rchi } 
        \delta^{n_\calP} \calP^\vectk
    }{ p^+_{\vect\rchi}!  n_{\calP}! } 
\nonumber \\
= \sum_{\mathclap{\substack{ \\ \\
        \left\{ n_{\vect\theta i} \right\},
        \left\{ p_{\vect\theta i} \right\},
        n_{\vect\rchi} 
        \text{  such that} \\
        \sum_{i=1}^{d_{\vect\theta}}  n_{\vect\theta i} p_{\vect\theta i}
        + n_{\vect\rchi} = n
     } } }
    \binom{ p^+_{\vect\theta} }{ p_{\vect\theta_1}, \dots, p_{\vect\theta d_\vect\theta } }
    (\frac{ \vectk\sast\delta^{n_{\vect\theta 1}} \Delta\vect\theta  }{n_{\vect\theta 1} !} )^{p_{\vect\theta 1} }
    \cdots 
    (\frac{ \vectk\sast\delta^{n_{\vect\theta d_{\vect\theta} }} \Delta\vect\theta }{n_{\vect\theta d_{\vect\theta} } !} )^{p_{\vect\theta d_{\vect\theta} } } 
    \cdddot_{(p^+_{\vect\theta})} 
    \frac{
    \partial_{\vect\theta}^{p^+_\vect\theta} 
    \delta^{n_\vect\rchi} \vect\rchi
    }{ p^+_{\vect\theta}!  n_{\vect\rchi}! } ,
\end{align}
\end{widetext}

\bibliography{main}

\end{document}